**Stability Analysis of an Inflatable Vacuum Chamber**

Sean A. Barton

5    Department of Physics,
Florida State University,
Tallahassee, FL  32306

**Abstract**

A light-weight "inflatable" tensioned-membrane-structure vacuum container is proposed and its
stability is analyzed.  The proposed structure consists of a pressurized lobed cylindrical "wall"
surrounding a central evacuated space.  Stability is analyzed by discretizing the system and
diagonalizing the second derivative of the potential energy.  The structure is found to be stable
15   when the pressure in the wall is greater than a critical pressure.  When membranes are non-
elastic, the critical pressure is found to be greater than the pressure required for equilibrium by a
factor of 4/3.  When membranes have only finite stiffness, a first order correction to the critical
pressure is found.  Preliminary experimental data shows that a stable structure can be made in
this way, and that the observed critical pressure is consistent with theory.  It is also found that
20   such structures can be designed to have net positive buoyancy in air.

**Introduction**

A structurally stable vacuum container that is of minimal total mass for a given evacuated
25   volume might have applications in airship design (buoyancy control) [1], aerospace (low
aerodynamic drag magnetic levitation launch systems) [2], industry (large industrial vacuum
chambers), transportation (supersonic maglev trains) [2], and solar energy production (solar
chimney technology) [3].  Unfortunately, issues of structural stability are often overwhelming in
the design of such a structure.

30

The history of lightweight vacuum containers is somewhat disconnected.  Von Guericke created
the first artificial vacuum around 1654 [4].  Traditional containers were thick heavy shells, the
thickness being required to give sufficient stability to prevent buckling.  In 1878, Tracy patented
an "aircraft" that aimed to derive lift from the buoyancy of a vacuum enclosed in an unstable
35   light-weight container [5].  In 1921, Armstrong patented another such craft that claimed to
stabilize its vacuum volume in an in-fact unstable inflated tensioned shell [6].  More recently,
Michaelis and Forbes have discussed the basic forces required to achieve equilibrium (not
stability) in a tensional vacuum vessel and have proposed the light-weight or weightless
inflatable vacuum chamber [7].  Lennon and Pellegrino have discussed the stability of inflated
40   structures [8] however a stability analysis of an inflated vacuum vessel (the purpose of the
current work) has not been carried out.

In the current work we propose an axially symmetric "cylindrical" structure composed of a
"wall" surrounding a central evacuated volume.  The wall consists of pressurized regions within
45   a network of tensioned membranes.  Rigorous stability analysis is carried out by (a) discretizing
the degrees of freedom of the system, (b) forming the matrix which represents the second derivative of the potential energy with respect to these degrees of freedom, and (c) diagonalizing
the matrix to confirm positive-definiteness and, hence, stability. The proposed structure is found
to be stable when sufficiently pressurized. Judicious choice of membrane materials and
50 pressurizing gas can lead to a structure that has over half of its total volume completely
evacuated and net positive buoyancy in air.

**The Proposed Structure**

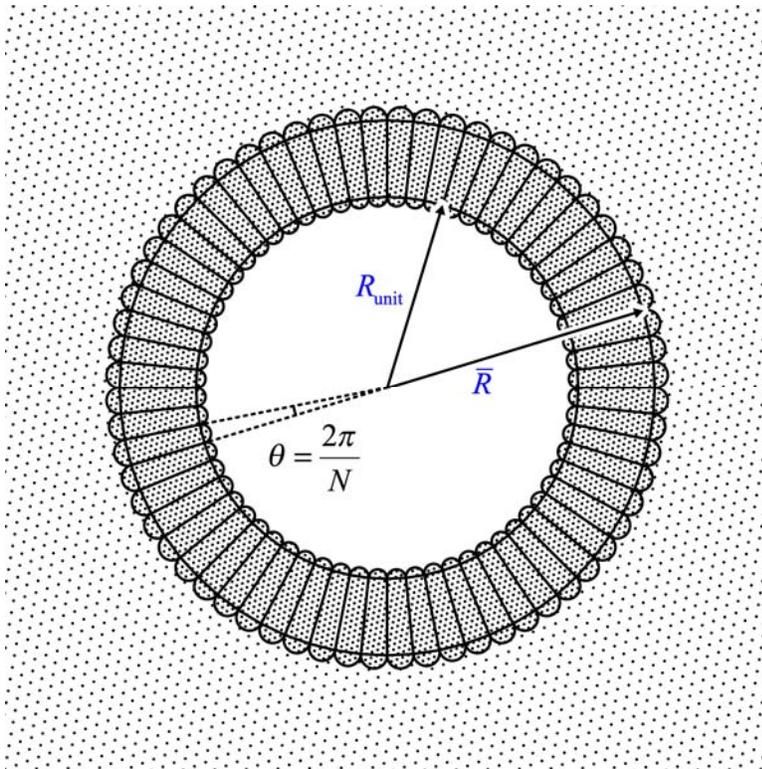

55
Figure 1, general cross-section of the proposed structure. The wall of the structure is composed
of membranes under tension (solid lines) containing pressurized gas (heavily hatched area). The
wall encloses the evacuated space at the center (unhatched area) isolating that space from the
ambient pressure (lightly hatched area). $R_{unit}$, $\overline{R}$, and $N$ are the inner radius, outer radius, and
60 number of sections respectively.

The proposed structure is shown and described in Figure 1 and the caption thereof. The
inspiration for the present design is as follows. The radial members must exist to transmit the
tension which will prevent the inner membrane from imploding. One can see this by considering
65 the mean stress tensor in the wall required to give equilibrium as indicated by the method of
sections. The hydrostatic pressure of the gas contributes positively and equally to all three
eigenvalues of the mean stress tensor, but the method of sections indicates (for any circular
cylinder subjected to hydrostatic pressure from inside or outside) that the eigenvalues of the
mean stress tensor in the wall must be in the approximate ratio 2 to 1 to 0 in the circumferential,
70 axial, and radial directions respectively. Thus (in addition to the hydrostatic pressure) there must
exist members under tension (the membranes) that contribute negatively to the eigenvalues
corresponding to radial and axial directions. Thus one adds membranes in the plane of the axial and radial directions to carry these tensions. The lobes are then added to terminate these tensions. If the curvature of the lobes is decreased (compared to Fig. 1), the tension in the lobes
75   is increased and somewhat redirected such that the eigenvalue of the mean stress tensor in the circumferential direction is reduced requiring additional pressure to maintain equilibrium. This is undesirable and thus the curvature in the lobes is kept at the maximum permitted by geometrical constraints. These radial members and lobes are sufficient to establish equilibrium; however, this geometry is highly unstable through what one might call the "accordion" effect,
80   similar to the instability of the hypothetical inflated lobed column described in the introduction of [8]. Thus the addition of the circumferential members are necessary to eliminate this instability.

Here we briefly consider some practical points related to the fabrication and use of such a
85   vacuum chamber. Likely, pressure will be supplied to a single compartment and inter-compartmental holes will allow pressure to distribute to all compartments. We note, for equilibrium, that the axial tension in each membrane is approximately one half the tension in the perpendicular direction. We also note that the vacuum chamber will require additional structures to close and seal the ends. The ends of the chamber may be capped with a single membrane in
90   the form of an concave hemisphere. It might also be capped with a complex network of membranes that extend the cylindrical wall into a convex hemispherical wall that closes the end. Where weight of the ends is of little concern, the end might be capped with a traditional compressive structure. We also note that the weight of the end structures as a fraction of the total weight is inversely proportional to the length and thus is negligible for long chambers. For
95   chambers where the length to diameter ratio is small, the end structures may enhance equilibrium and stability; in our analysis however we will consider the situation where the end-structures are very far from the section under analysis.

### A Note about Units

100

Before analyzing the proposed structure, we first define our notational convention. Each physical quantity with appropriate units is represented by a barred symbol (i.e. $\bar{R}$ ); each corresponding "reduced" quantity (a dimensionless number which is the physical quantity divided by a reference value) is denoted by the unbarred symbol (i.e. $R$). Many of the equations
105   that follow are more conveniently expressed in terms of these dimensionless quantities. The reference values for the physical parameters are indicated with a subscript "unit" (i.e. $R_{unit}$ ) and are given in Appendix A. For example, the reduced values of tension, potential energy, and outside radius are

110   $$T \equiv \frac{\bar{T}}{T_{unit}}, \ U \equiv \frac{\bar{U}}{U_{unit}}, \ \text{and} \ R \equiv \frac{\bar{R}}{R_{unit}}$$

respectively. In the text we will refer to the physical quantities and the reduced quantities interchangeably. For example, we will refer to both $\bar{U}$ and $U$ as "potential energy". Which meaning is intended will be clear from the context.

115

### Modeling of the Proposed Structure

We wish to analyze the stability of the structure depicted in Figure 1. The structure is axially symmetric and of uniform cross-section, i.e. invariant under translations in the direction perpendicular to the plane of the drawing. We will analyze the most general form of this system having $N$ sections ($N$=64 in Fig. 1) and having an outside tension hoop of vertex radius $R$ (reduced) where the vertex radius of the inside tension hoop is taken to be the reference length $R_{unit}$. ( $R \equiv \bar{R}/R_{unit} \cong 1.4$ in Fig. 1.) For structural considerations, the central vacuum will be assumed to be "complete" (absolute pressure of exactly zero atmospheres). The absolute pressure within the wall is $P$ (where the reference pressure $P_{unit}$ is the ambient pressure). (Recall $P \equiv \bar{P}/P_{unit}$ .)

In order to analyze the equilibrium and stability of the system, we must write its potential energy $U$ as a function of configuration or deformation. We characterize the configuration of the system by coordinates $x_{ni}$ specifying the radial and circumferential displacements of the $N$ inside vertices and the $N$ outside vertices according to the convention established in Figure 2. Notice in Figure 2 that the first subscript specifies which unit cell and that the second subscript specifies which degree of freedom within the unit cell. Thus $x_{ni} = 0$ characterizes the nominal, intended, or undeformed configuration, which corresponds to an extremum (local minimum, local maximum, or saddle point) in the total potential energy for any given value of $P$. One then writes $U$ as a function of these $x_{ni}$. For equilibrium (or instantaneous balance of forces), one need only confirm that the first derivatives of $U$ with respect to the $x_{ni}$ are all zero. For stability, the second derivative of $U$ with respect to any and all linear combinations of the $x_{ni}$ must be non-negative. Third and higher-order derivatives are neglected in stability analysis as displacements are assumed to be small. For this reason our representation of $U$ need be valid only to second order about the nominal configuration.

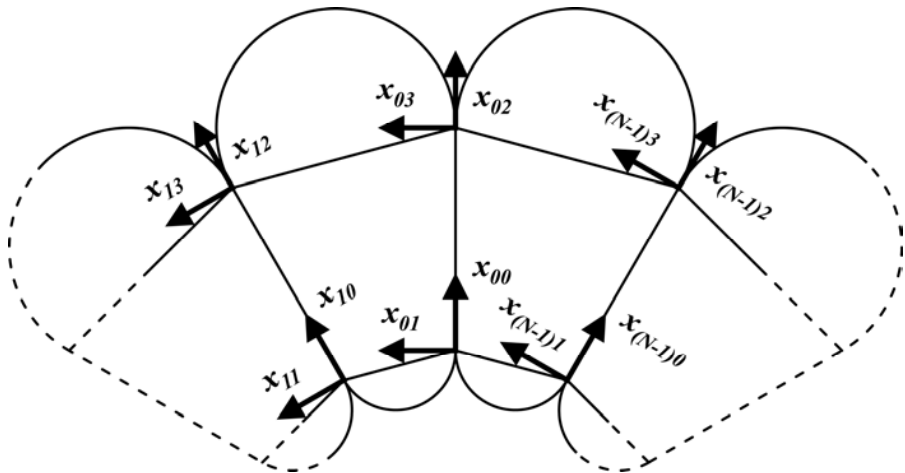

Figure 2, definition of the 4$N$ degrees of freedom of the structure $x_{ni}$

In order to write an expression for $U$, we consider the forms of potential energy that the system possesses. The system has two types of potential energy, "solid-elastic" energy and "pressure-volume" energy. Each tensioned membrane has a solid-elastic energy of the form

$$\Delta U_{\text{membrane}} = T \cdot \Delta l + \frac{1}{2} K \cdot \Delta l^2 \tag{1}$$

150    where $\Delta l$ is the change in length of the membrane relative to the equilibrium configuration, $T$ is the pretensioning, and $K$ is the elastic constant. This equation is valid only to second order in $\Delta l$ and thus the representation of $\Delta l$ need be valid only to second order in the $x_{ni}$. To achieve equilibrium, $T$ will be different for different membranes and since the elastic constant of a membrane $K$ depends on its length, thickness, and elastic modulus, $K$ will also be different for
155    different membranes.

Each volume under pressure has a pressure-volume energy of the form

$$\Delta U_{\text{gas}} = -\Delta p \cdot \Delta V \tag{2}$$

where $\Delta p$ is the difference in pressure across the boundary defining the volume and $\Delta V$ is the
160    change in volume relative to the equilibrium configuration. In order for this equation to be valid to second order in $\Delta V$, one must make the simplifying assumption that the pressure $P$ is constant during any change in volume. Thus, we assume we are connected to a large reservoir that maintains constant pressure. If this is not true, pressure can increase with a decrease in volume thus enhancing stability. This assumption can lead to a false conclusion that the system is
165    unstable but can never lead to a false conclusion that the system is stable. We call this a "failure-safe" assumption. As with $\Delta l$, the representation of $\Delta V$ must be valid to second order in the $x_{ni}$.

To simplify the form of $U$, we note that all pairs of circumferential nearest-neighbor vertices are connected by a curved membrane and a straight membrane that enclose a pseudo-semicylindrical
170    volume. The solid-elastic energy of these two membranes and the pressure-volume energy of the enclosed space are all determined only by the distance between the pair of vertices. Thus there is no need to represent these energies separately in $U$ as they can be absorbed into a single hypothetical "spring" with an "effective" pretensioning $T$ and an "effective" elastic constant $K$ (see $a$, $\alpha$, $b$, and $\beta$ defined in the caption of Figure 4).
175

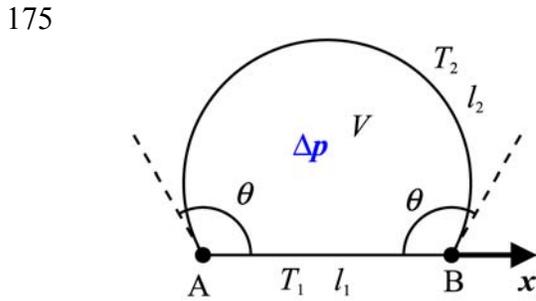

Figure 3, subsystem for illustrating the meaning of "effective" tension

As an example of an "effective" tension, take the subsystem of Fig. 3 consisting of two nodes,
180    two membranes, and the enclosed pressurized volume. (Here, in this example, the symbols $\theta$ and $x$ are unrelated to previous uses of the same symbols.) We can write the potential energy of the subsystem as a power series in the horizontal displacement of point B, $x$.

$$U = U_0 + xU_1 + \frac{1}{2}x^2 U_2 + \frac{1}{3!}x^3 U_3 + \dots \quad \text{where} \quad U_1 = \frac{dl_1}{dx}T_1 + \frac{dl_2}{dx}T_2 - \frac{dV}{dx}\Delta p = T_1 + T_2 \cos\theta \, .$$

185

We now identify $U_1$ as the effective tension between point A and point B and we see that it can be calculated from the membrane tensions ($T_1$ and $T_2$). (Notice, if $\theta$ is greater than 90° and $T_2$ is sufficiently large compared to $T_1$, that the effective tension $U_1$ can be negative.)

190   Again to simplify calculation, we assume that there is an infinitesimal clearance angle between the curved membranes so that we need not consider their interference with each other. If they were to interfere, stability would be enhanced as this is an additional constraint on the system. Again this is a "failure-safe" assumption.

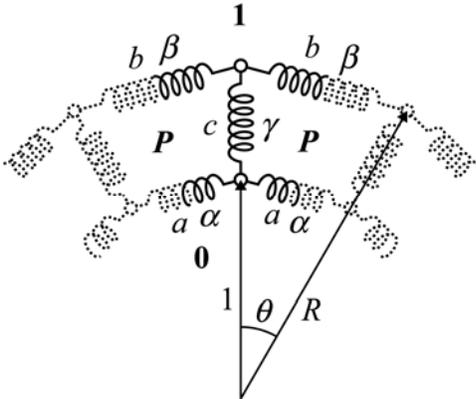

195   Figure 4, one "unit cell" of the idealized model of the system showing the pretension $c$, the spring constant $\gamma$, the "effective" pretensions $a$ and $b$, the "effective" spring constants $\alpha$ and $\beta$, and pressures in bold type

**Stability Analysis**

200   The $\Delta l$'s and the $\Delta V$'s are written as polynomials in the $x_{ni}$ retaining terms up to second order. These polynomials are inserted into the general forms of $\Delta U_{\text{membrane}}$ and $\Delta U_{\text{gas}}$ to give $U$ which we write as a power series in $x_{ni}$,

205

$$U = U_0 - \sum_{ni} F_{ni} x_{ni} + \sum_{nimj} \tfrac{1}{2} K_{nimj} x_{ni} x_{mj} + \cdots \qquad (3)$$

where $U_0 \equiv U|_{x=0}$, $-F_{ni} \equiv \dfrac{\partial U}{\partial x_{ni}}\bigg|_{x=0}$, and $K_{nimj} \equiv \dfrac{\partial^2 U}{\partial x_{ni} \partial x_{mj}}\bigg|_{x=0}$ (necessarily symmetric). (Here the symbol $K_{nimj}$ is unrelated to the previously used symbol $K$ in (1).) In (3), $K_{nimj}$ can be thought of

210   as a "block" matrix [9] (p. 8) and $F_{ni}$ and $x_{ni}$ can be thought of as "block" vectors. We are able to find all of the elements of $F_{ni}$ and $K_{nimj}$ by considering the contributions to (3) from all of the hypothetical springs (1) and the pressurized volumes (2) of Figure 4 (see appendices B and C for example contributions). Because potential energy is relative, we are free to set $U_0 = 0$. For equilibrium (all $\ddot{x}_{ni} = 0$) we require $F_{ni} = 0$. It can be shown (see appendices B and C) that

215   $F_{n0} = SC - 4\delta^2 A - 2\varepsilon P$, $F_{n2} = -SC - 4\delta^2 RB + 2\varepsilon RQ$, and $F_{n1} = F_{n3} = 0$ where (for notational

convenience) we have defined $S \equiv R - 1$, $C \equiv \dfrac{c}{S}$, $A \equiv \dfrac{a}{2\delta}$, $B \equiv \dfrac{b}{2\delta R}$, $Q \equiv P - 1$, $D \equiv \cos\dfrac{\theta}{2}$,

$\delta \equiv \sin\dfrac{\theta}{2}$, $E \equiv \dfrac{\cos\theta}{2}$, and $\varepsilon \equiv \dfrac{\sin\theta}{2}$ as in section "Nomenclature" below. Notice that the $A$, $B$, and $C$ terms originate from the membranes and the $P$ and $Q$ terms originate from the pressurized gas. These conditions lead to $A = \dfrac{SC - 2\varepsilon P}{4\delta^2}$ and $B = \dfrac{-SC + 2\varepsilon RQ}{4\delta^2 R}$ but do not lead to a unique

220　　solution for $A$, $B$, and $C$. By the method of sections, we find that the tension in a membrane $T$, the radius of cylindrical curvature of the membrane $r$, and the pressure difference across the membrane $\Delta p$ are related by $\Delta p \cdot r = T$. This determines the tension in the curved membranes of the outer and inner lobes. Then requiring that the tension in the tension hoops be non-negative, we find a constraint on the tension in the radial membranes,

225　　$$\frac{2\delta P}{DS} \leq C \leq \frac{2\delta RQ}{DS} \tag{4}$$

and since the LHS of (4) must be less than the RHS, we find

$$P \geq \frac{R}{R-1} \tag{5}$$

as shown by Michaelis and Forbes [7].

230　　It can now be seen (given the spectrum of solutions for $A$, $B$, and $C$) that the system is statically-indeterminate. The actual values of $A$, $B$, and $C$ will (in practice) depend on the precise unstressed lengths of the membranes. Slight variations in these unstressed lengths will determine the distribution of forces (and $C$) when the load (pressure) is applied. The value of $C$ may be difficult to control without very precise means of manufacture, however we will assume that the

235　　structure can be fabricated with enough precision that $C$ can be made to fall within the range required for equilibrium (4). In practice, one may test the structure through inflation to determine $C$. If $C$ is found to lie outside the desired range, the inner or outer tension hoops may be lengthened or shortened to adjust $C$. One may note that the minimum value of $C$ (4) corresponds to zero tension in the inner tension hoop, the maximum value of $C$ (4) corresponds to zero tension in the outer tension hoop, and the minimum value of $P$ (5) corresponds to zero

240　　tension in both the inner and outer tension hoops.

Note that, while it is true that the pretensioning in a membrane cannot be negative, no such restriction applies to an effective tension (as illustrated in the example of Figure 3); for example,

245　　the effective tension $b$ is often negative.

Returning to the stability analysis, we wish to explore only infinitesimal deformations about the equilibrium position and thus terms third order in $x_{ni}$ are negligible compared to the second order terms. Thus we write simply $U = \sum_{nimj} \frac{1}{2} K_{nimj} x_{ni} x_{mj}$ or equivalently

250

$$U = \frac{1}{2} \sum_{nimj} x_{ni}^{*} K_{nimj} x_{mj} \tag{6}$$

as we are free to do because the $x_{ni}$ are real. Recall that $K_{nimj}$ is symmetric and real and thus Hermitian and therefore has orthogonal eigenvectors [9] (p. 268). For stability, $K_{nimj}$ must be such that no real $x_{ni}$ leads to a $U$ that is less than zero. To determine if $K_{nimj}$ is of such form, we wish to make unitary transformation to a new basis $\chi_{k\mu} = \sum_{k\mu ni} X_{k\mu ni} x_{ni}$ where

$$U = \tfrac{1}{2} \sum_{k\mu l\nu} \chi_{k\mu}^* \kappa_{k\mu l\nu} \chi_{l\nu} \,, \tag{7}$$

such that $\kappa_{k\mu l\nu} = \delta_{kl} \delta_{\mu\nu} \kappa_{k\mu} = \sum_{nimj} X_{k\mu ni} K_{nimj} X_{mjl\nu}^\dagger$ where the star "*" denotes complex conjugate, the dagger denotes Hermitian conjugate (e.g. $X_{mjl\nu}^\dagger = X_{l\nu mj}^*$), and $\delta_{kl}$ is the Kronecker delta symbol (i.e. the matrix elements of the identity matrix). Thus we have made a unitary similarity transformation to the diagonal representation of $K_{nimj}$, where the $X_{k\mu ni}$ (which define the similarity transform) are the orthogonal eigenvectors of $K_{nimj}$ with the eigenvalues $\kappa_{k\mu}$. The condition of stability is satisfied when the $\kappa_{k\mu}$ are non-negative. Thus our problem reduces simply to confirming that $K_{nimj}$ is positive semidefinite (or non-negative definite [9] (p. 7)).

### Confirming that $K_{nimj}$ is Positive-Semidefinite

Beginning with $K_{nimj}$, we make unitary similarity transformations with the goal of eventually finding the diagonal representation of $K_{nimj}$. Because the system has rotational symmetry and couplings only between nearest-neighbor unit cells, we can write $K_{nimj} = \delta_{nm} G_{ij} + \delta_{(n+1)m} J_{ij} + \delta_{(n-1)m} J_{ij}^\top$ where $G$ describes couplings within a unit cell and $J$ describes couplings between neighboring unit cells. (Given the cylindrical symmetry, the Kronecker delta symbol is understood to function cyclically (e.g. $\delta_{0N} = 1$).) With the intent to diagonalize $K_{nimj}$ we notice the symmetry $K_{nimj} = K_{(n+1)i(m+1)j}$ and thus $K_{nimj}$ commutes with the operation of rotating the entire system by one unit cell, $\hat{\Phi}$ (the matrix elements of which are $\hat{\Phi}_{nimj} \equiv \delta_{(n+1)m} \delta_{ij}$). And because matrices that commute can be simultaneously diagonalized, each eigenvector of $K_{nimj}$ must lie completely within a subspace spanned by the degenerate eigenvectors of $\hat{\Phi}$ characterized by a single eigenvalue. Thus if we transform to a basis of the eigenvectors of $\hat{\Phi}$, $K_{nimj}$ becomes "block" diagonal in blocks corresponding to the distinct eigenvalues of $\hat{\Phi}$. We note that the eigenvectors of any translation operator are Fourier components. Thus following a technique similar to that used in [10], we block diagonalize $K_{nimj}$ with a Fourier transform $V_{kn}$ to give

$$K'_{kilj} = \sum_{nm} V_{kn} K_{nimj} V_{ml}^\dagger \tag{8}$$

where $V_{kn} \equiv \dfrac{e^{-ikn\theta}}{\sqrt{N}}$.  We find

$$K'_{kilj} = \sum_{nm} \frac{e^{i(lm-kn)\theta}}{N}(\delta_{nm}G_{ij} + \delta_{(n+1)m}J_{ij} + \delta_{(n-1)m}J^{\mathrm{T}}_{ij}) = \delta_{kl}(G_{ij} + e^{ik\theta}J_{ij} + e^{-ik\theta}J^{\mathrm{T}}_{ij}) \equiv \delta_{kl}K'_{kij} \qquad (9)$$

and thus the diagonal blocks of $K'_{kilj}$ are

290

$$K'_{kij} = G_{ij} + e^{ik\theta}J_{ij} + e^{-ik\theta}J^{\mathrm{T}}_{ij}. \qquad (10)$$

We again realize that there must exist an additional transformation $W_{k\mu i}$ such that

$$\sum_{ij} W_{k\mu i}K'_{kij}W^{\dagger}_{kj\nu} = \delta_{\mu\nu}\kappa_{k\mu} \quad \text{(i.e. } W_{k\mu i}V_{kn} = X_{k\mu ni}) \text{ where } W^{\dagger}_{kj\nu} = W^{*}_{k\nu j} \text{ and thus the } W_{k\mu i} \text{ are the}$$

eigenvectors of $K'_{kij}$ with eigenvalues $\kappa_{k\mu}$.  Thus, our problem further simplifies to confirming

295    that each $K'_{kij}$ has no negative eigenvalues.  By considering all contributions to $U$ like $\Delta U_{\mathrm{membrane}}$
(1) and $\Delta U_{\mathrm{gas}}$ (2) (see Appendices B and C respectively for examples), one can determine the
matrix elements of $G_{ij}$ and $J_{ij}$.  For example $G_{00}$ is the self-coupling elastic constant for any
inside node moving in the radial direction.  This takes major contributions from the elastic
constant of the radial membrane $\gamma$ and the effective tension in the inside tension hoop $a$.  It can
300    be shown that $G_{00} = \gamma + 2D^2A + 2\delta^2\alpha$ , $G_{11} = C + 2\delta^2 A + 2D^2\alpha$ , $G_{22} = \gamma + 2D^2B + 2\delta^2\beta$ ,
$G_{33} = C + 2\delta^2 B + 2D^2\beta$ , $G_{02} = G_{20} = -\gamma$ , $G_{13} = G_{31} = -C$ , and all other elements in $G_{ij}$ equal
zero, and that $J_{00} = -D^2A + \delta^2\alpha + EP$ , $J_{11} = \delta^2 A - D^2\alpha + \varepsilon P$ , $J_{22} = -D^2B + \delta^2\beta - \varepsilon Q$ ,
$J_{33} = \delta^2B - D^2B - \varepsilon Q$ , $J_{01} = -J_{10} = \varepsilon A + \varepsilon\alpha + EP$ , $J_{23} = -J_{32} = \varepsilon B + \varepsilon\beta - EQ$ , and all other
elements in $J_{ij}$ equal zero.

305

To aid in confirming the positive definiteness of each $K'_{kij}$, we will assume that all of the elastic
constants ($\alpha$, $\beta$, and $\gamma$) are large compared to the other variables ($A$, $B$, $C$, $P$, and $Q$).  This
approximation is often valid for inflatable structures because the "effective" elastic modulus (or
Young's modulus) of an ideal diatomic gas is only 1.4 times its "effective" yield strength (or
310    pressure).  This is in contrast to solids which often have elastic moduli several orders of
magnitude larger than their yield strengths.  Thus compared to gases, solids are "stiff".  The same
approximation was made by Lennon and Pellegrino in their analysis [8].  We will call this the
"stiff solid" approximation.  We will later reexamine this approximation to find a first-order
correction.  We know that $\alpha$, $\beta$, and $\gamma$ must give positive contributions to the eigenvalues as they
315    represent springs with only positive spring constants.  Thus, in the limit that they are large, the
only possibility of finding a negative eigenvalue will be to look in the null space of the $\alpha$, $\beta$, and
$\gamma$ terms.

Now neglecting $A$, $B$, $C$, $P$, and $Q$ terms and considering only $\alpha$, $\beta$, and $\gamma$ terms in $K'_{kij}$, we have
320    $K''_{kij}$.  A simple analysis indicates that $K''_{kij}$ has exactly one null vector (unnormalized)

$$\left[ -iD\sin\frac{k\theta}{2} \quad \delta\cos\frac{k\theta}{2} \quad -iD\sin\frac{k\theta}{2} \quad \delta\cos\frac{k\theta}{2} \right]$$

for each $k$ except $k=0$.  When $k=0$ one finds the two null vectors,

$\begin{bmatrix} 0 & 1 & 0 & 0 \end{bmatrix}$ and $\begin{bmatrix} 0 & 0 & 0 & 1 \end{bmatrix}$.

325     Thus we have found two potentially unstable modes for $k$=0 and one for every other $k$ for a total of $N$+1 modes allowed within the "stiff solid" approximation.  The remaining $3N$-1 modes have eigenvalues going to positive infinity in the "stiff solid" approximation and are thus stable and not of interest.  We are interested only in the $N$+1 modes in which the elastic constants $\alpha$, $\beta$, and $\gamma$ do not contribute to the eigenvalue and thus the stability is governed by the pressure differences.

330     It is in this $N$+1 dimensional space that we expect to find the $N$+1 non-infinite energy eigenmodes of the system.  For $k$=0 which allows more than one mode, we must again diagonalize in that two-dimensional subspace to find the eigenmodes.  We thus operate the $K'_{kij}$ (including the $A$, $B$, $C$, $P$, and $Q$ terms) onto these null eigenvectors of $K''_{kij}$ to determine if their eigenvalues (considering all terms) are positive or negative.  One need not normalize the vectors

335 in order to simply determine the sign of the eigenvalue.

    In the basis of the two $k$=0 modes allowed in the "stiff solid" approximation, the matrix elements of $K'_{kij}$ are

$$\begin{bmatrix} RC & -C \\ -C & R^{-1}C \end{bmatrix}.$$

340     The determinant of this matrix is found to be zero and the trace is found to be positive indicating one zero eigenvalue and one positive eigenvalue.  (The zero eigenvalue corresponds to overall rotation of the system.)  We continue with the remaining $N$-1 $k$ values in search of the mode of greatest instability.  By operating $K'_{kij}$ on the remaining null eigenvectors of $K''_{kij}$ (where $k$ does not equal 0) we find that the sign of each eigenvalue is determined by the sign of

345 $$\left(\cos\theta - \cos(k\theta)\right)\left(CS^2\cos\theta - R\sin\theta - CS^2\cos k\theta + R\cos k\theta\sin\theta\right).$$

    Notice that when $k$=1 or $N$-1 we have an eigenvalue of zero independent of $C$, $R$, or $\theta$.  Linear combinations of these two modes correspond to overall translation of the system along global "x" and "y" axes.

350     For $2 \leq k \leq N-2$ notice that the sign of each eigenvalue is determined by the sign of

$$\frac{CS^2}{R} - \frac{(1-\cos k\theta)\sin\theta}{\cos\theta - \cos k\theta}.$$

Taking first and second derivatives of this with respect to $k$, it is quickly found that it is most negative when $k$=2 or $N$-2.  And so the greatest possibility of making the eigenvalue less than zero is for the modes $k$=2 and $k$=$N$-2.  Hence these are always the most unstable modes and thus

355 they determine the overall stability of the system.

    Notice that the condition for overall stability of the system is thus

$$\frac{CS^2}{R} \geq \frac{(1-\cos 2\theta)\sin\theta}{\cos\theta - \cos 2\theta}. \tag{11}$$

    From this it is seen that a maximum $C$ enhances stability; thus reducing tension in the outer

360 tension hoop is desirable.  However in keeping with our "failure-safe" assumptions, we will assume that $C$ takes its minimum value (4).  This leads to an expression for stability in terms of pressure and radius,

$$P\frac{S}{R} \geq \frac{4\left(\cos\frac{1}{2}\theta\right)^4}{1+2\cos\theta}. \tag{12}$$

The right side of eq. (12) goes to 4/3 in the small $\theta$ limit (many sections or large $N$), and does not
exceed 4/3 for any reasonable $N$ (larger than 3). Thus, the "failure-safe" requirement for overall
stability of the entire system given the "stiff solid" approximation is

$$\boxed{P \geq \frac{4}{3}\frac{R}{R-1}.} \tag{13}$$

Notice that the requirement for stability (13) compared to that of equilibrium (5) is to simply
increase the absolute pressure by a factor of 4/3.

One may be surprised that the result (13) is somewhat insensitive to the precise value of $N$.
Recall however that some $N$ dependence has been removed between equations (12) and (13).
For example, if we take (12) with $N$=4 ($\theta$=π/2), the factor of 4/3 in (13) becomes unity. In this
case, however, the external and internal lobes occupy a large fraction of the total volume of the
system and thus $N$=4 is not practical. To minimize the total volume of the system while
maximizing the evacuated volume, large values of $N$ are of interest. In this limit, the right-hand
side of (12) tends to 4/3 and never exceeds 4/3, thus (13) is a general result that is "failure-safe".
In the large-$N$ limit (given the stiff-solid approximation), the complete potential energy of the
system can be approximated as the pressure-volume energy of a system of two coaxial thin-
walled cylinders (of radii unity and $R$) constrained such that each point on the inner cylinder
remains a fixed distance $S \equiv R-1$ from the point on the outer cylinder that corresponds when the
cylinders take the nominal circular cross-section. This pressure-volume energy would depend on
the overall distortion of the cylinders, not on the roughness of the boundaries which each
cylinder defines. Thus when $N$ is large, the system can be approximated by a continuous system
for which the prefactor is exactly 4/3.

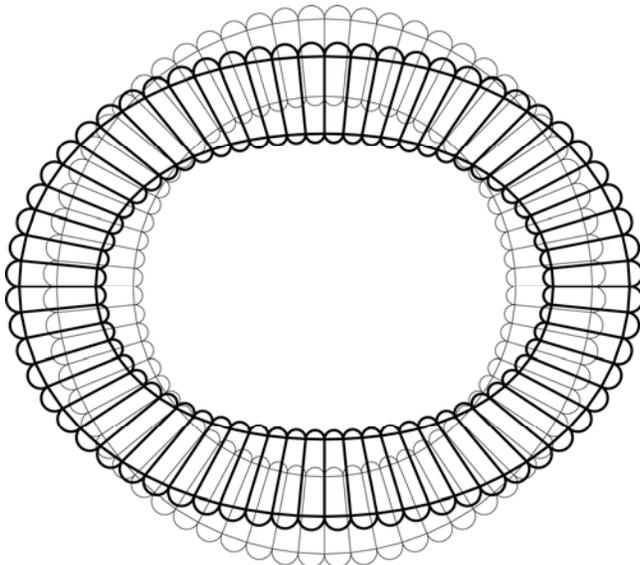

Figure 5, linear combination of $k$=2 mode and $k$=$N$-2 mode

Given a violation of the stability requirement (13), failure by the $k$=2 and $k$=$N$-2 modes may occur. It is instructive to discuss these failures in direct physical terms. A linear combination of these two modes is shown in Figure 5. Notice that the intended cylindrical form has distorted to give a pseudo-elliptical cross section. Other linear combinations of the two failure modes give the same elliptical distortion but with the major axis of the ellipse oriented differently. From Figure 5, one can see that the deformation is allowed because, while the wall structure is bending-stiff, it is shear-weak. The shear is concentrated at the quarter-points approximately 45° from the major and minor axes of the ellipse. This shear-weakness (in contrast to the bending-stiffness) can be easily understood given the "stiff solid" approximation. This approximation assumes that no elastic energy can be stored in the membranes and hence the length of every membrane is fixed. With no means of stretching any membrane, one can show that the only way to accommodate bending of the wall structure (without shear) is to eliminate the tension in either the outer or inner circumferential membranes such that they buckle. This leads to an increase in potential energy of the wall because the pressurized volume is then reduced. One can show that when the wall is bent by this mechanism, the increase in potential energy is proportional to the absolute value of the change in the wall's curvature; and because the absolute value function has an infinite second derivative at its origin, this bending mode is "infinitely" stable (given the stiff-solid approximation). A similar consideration of the shearing mode illuminates the origin of the weakness. To accommodate shearing of the wall structure, there is no need to change the length of any membrane. Like the bending, the shearing causes a reduction in the pressurized volume, but one can show that this change in volume is proportional to the square of the change in curvature, in contrast to the bending mode where the change in volume is proportional to the absolute value of the change in curvature. Because the potential energy in this case has a finite second derivative, the shearing mode is much more active than the bending mode.

**Stability Without the Stiff-Solid Approximation**

The analysis thus far depends on the "stiff solid" approximation. Let us now reexamine that approximation. If one assumes that all solid components of a system have infinite stiffness (as the "stiff-solid" approximation assumes), then any continuous solid system is stable and there is no need to consider inflatable structures to enhance stability. However, experience shows that as the wall of a vacuum chamber becomes progressively thinner, the stiff-solid approximation at some point becomes inadequate. We will now show that the same is true of inflatable vacuum chambers, that equation (13) is inadequate when $R$ is sufficiently small. To understand this, we must go back and find the critical $P$ that makes $K'_{2ij}$ positive-semidefinite without assuming that $\alpha$, $\beta$, and $\gamma$ are large. To arrive at the smallest possible values for $\alpha$, $\beta$, and $\gamma$ (and thereby achieve a failure-safe result), we will assume that every membrane is only thick enough to just meet its strength requirement. (We continue to take $C$ at its minimum value.) We further assume that membranes comprising the inner and outer tension hoops are of equal thickness and only of sufficient strength to withstand venting of the chamber. We also neglect the small contributions that the curved membranes make to the effective elastic constants $\alpha$ and $\beta$. Given the unitless number $M$ which is the elastic modulus divided by the tensile strength, these assumptions determine the elastic constant of every membrane. We insert these explicit forms of $\alpha$, $\beta$, and $\gamma$ into $K'_{2ij}$. We will still assume that $\theta$ is small (a "failure-safe" assumption) and thus expand each element of $K'_{2ij}$ in a power series in $\theta$ keeping only the leading non-zero term. For every non-zero element this is the term first-order in $\theta$. We then explicitly set the determinant to zero,

$$M^3 \frac{3Q^2 P(3PS - 4R)S\theta^4}{R^2} +$$

440

$$M^2 \frac{-3QP\left(3P^2S^4 + S^2R(3R-4) - P\left(3 - 18R + 27R^2 - 22R^3 + 6R^4\right)\right)\theta^4}{R^2S} +$$ (14)

$$M \frac{-18QP^2(PS - R)S\theta^4}{R^2} + \frac{12P^3\theta^4}{RS} = 0,$$

to find the critical $P$. (Note, the elastic constant of each membrane depends on its thickness, and the thickness chosen depends on the yield strength. Thus (14) is not dependent upon elastic

445 modulus alone, but upon the ratio $M$.) If we take the limit that $M$ is large by only considering the term third-order in $M$, we find that this determinant equals zero when $P = P_0 \equiv \frac{4}{3}\frac{R}{R-1}$ as expected (13). To find the first order correction to $P$ when $M$ is not so large, we consider the third- and second-order terms. We expand this in a power series in $P - P_0 \equiv \Delta P$ keeping only zero- and first-order terms in $\Delta P$. Then solving this for $\Delta P$ and looking at the term first-order in

450 $M^{-1}$, we find that

$$\frac{\Delta P}{P_0} \approx \frac{1}{M} \frac{R(11 + 3R + R^2 + R^3)}{4(R-1)^2(R+3)}$$ (15)

which blows up as $R$ goes to unity. In this limit we find

455

$$\frac{\Delta P}{P_0} \approx \frac{1}{(R-1)^2 M}.$$ (16)

(For an example case where $R$=1.4 and $M$=65 (typical for carbon fiber), we find that $\Delta P/P_0 \approx 15.2\%$ (15). A numerical solution of eq. (14) in the same case gives $\Delta P/P_0$=14.8%.) In

460 summary, by comparing $\Delta P$ with $P_0$ (16) we find that (13) is inadequate when $R$ is close to or less than $1 + M^{-1/2}$.

**Preliminary Experimental Data**

465 A model was constructed of polyester film bonded with acrylic adhesive having $R$=2 and $N$=12 (see Figure 6). The pressure in the wall of the structure was raised above atmospheric pressure by 1.01±0.10 psi (6.96±0.69 kPa). Partial evacuation of the central volume was then begun. Stability was maintained until the central pressure reached 0.72±0.10 psi (4.96±0.69 kPa) below atmospheric pressure. At this point the central volume began to decrease by means of the $k$=2

470 deformation, thus preventing further reduction in pressure even though pumping continued. Pressure in the wall of the structure was then raised to 2.0 psi (13.8 kPa) above atmospheric pressure, however before another measurement could be made, an internal failure of the acrylic adhesive redistributed stresses to rupture the exterior film and hence the pressure was lost. The model was not repaired.

475

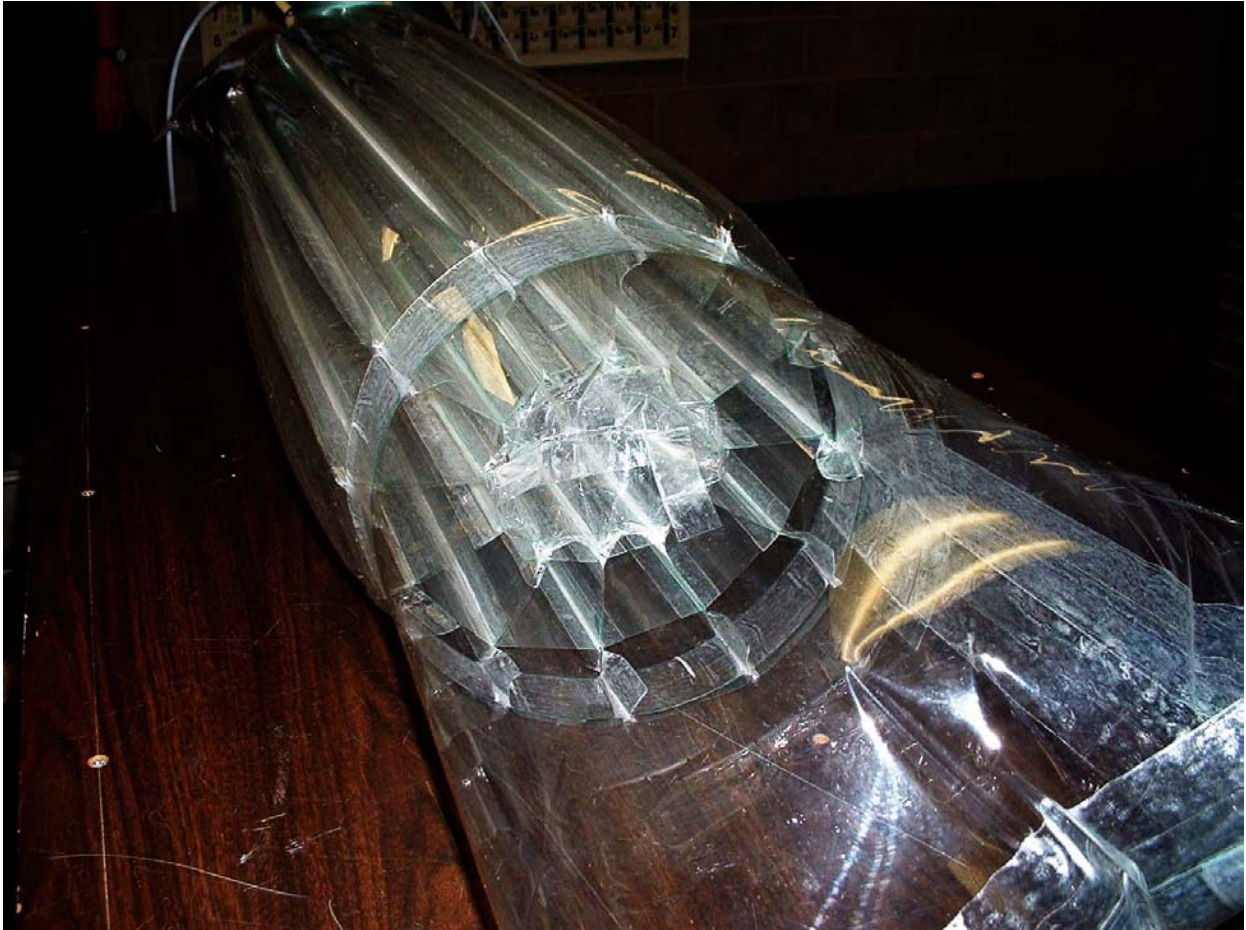

Figure 6, experimental model of the inflatable vacuum chamber

The current stability theory is not directly applicable to this experiment because the absolute
480 pressure in the central space is not zero, however notice that (1) and (2) are invariant under a global offset in the hydrostatic pressure. Thus the current experiment can be analyzed by subtracting from all pressures, the pressure in the central space. Therefore the one data point obtained has the same stability as $P \equiv \dfrac{\overline{P}}{P_{\text{unit}}} = \dfrac{1.01 \text{ psi} + 0.72 \text{ psi}}{0.72 \text{ psi}} = 2.40$. Given $N$=12 and $R$=2

and knowing the bounds on $C$ (4), the theory predicts the critical $P$ required for stability between
485 2.27 and 2.55 (11) which is consistent with this observation.

**Application of Theory**

To apply this theory to achieve a structure that is lighter than air, careful consideration is
490 required. A simple analysis shows that inflatable vacuum chambers of a cylindrical form contain at least twice as many moles of pressurized gas as the moles of gas they displace [7]. Thus pressurizing with air can never lead to a structure that is lighter-than-air. Additionally one should note that, if a particular design is found to possess sufficient strength to transmit the required forces when evacuated, these forces can increase when the chamber is vented (especially in the tension hoops) leading to failure of the system unless $P$ is simultaneously reduced. However, if pressurization is accomplished with helium, and membranes are constructed of advanced materials such as Kevlar- or carbon-fiber composites, calculations show that such a structure could be lighter-than-air with over half of its volume completely evacuated and still resist failure when vented. For example, if $R \approx 1.4$ and the structure is constructed of 60% carbon fiber/40% polyester composite with a safety factor of unity and pressurized with helium, it will have a total mass about one third the mass of air it displaces and (with sufficiently large $N$) over half of its total volume is completely evacuated.

If net positive buoyancy is not necessary, more basic materials and pressurized air could be used while still reducing the required amount of raw material by about two orders of magnitude compared to conventional chamber designs.

**Future Directions**

Preliminary analysis suggests that the structure of Figure 1 can be modified to further enhance stability and ease of fabrication. While all of the modified structures shown in Fig. 7 might have greater stability, the greatest stability is likely to be found in the structure of Fig. 7(a) or 7(b). The structure of Fig. 7(c), however, appears to be easiest to fabricate.

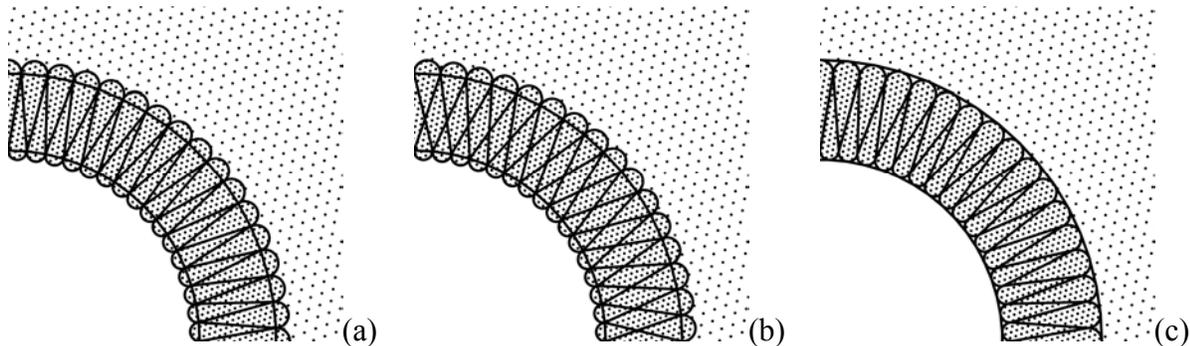

Figure 7, three possible modifications of the structure of Fig. 1

The logic behind the modification of Fig. 1 to arrive at Fig. 7(a) is that when the radial membranes are slightly diagonal, the shearing mode (the mode of failure) may be somewhat more stable. Because of this modification, the circumferential eigenvalue of the mean stress tensor in the wall may be additionally reduced and thus the pressure required for equilibrium might be slightly greater (which is of no consequence unless this is the mode of failure). Thus setting the angle of the radial membranes slightly off the exact radial direction could reduce the critical pressure; however, an excessive angle could result in the pressure required for equilibrium becoming larger than the pressure required for stability. Thus it appears that there is an optimal angle given a particular situation. Of course the angle is related to $N$ and $R$, however given $N$ and $R$, there still remains some freedom in the choice of this angle as one can have the membranes cross each other or reflect off of each other at one (or more) points intermediate their length as in Fig. 7(b).

The modification of Fig. 7(a) to arrive at Fig. 7(c) is simply to increase ease of fabrication. All members still serve the same purposes (lobes to terminate radial tensions and circumferential members to prevent the "accordion" effect), but the rearranged geometry now permits construction from three continuous membranes (a first membrane on the inner surface, a second membrane on the outer surface, and a third membrane meandering between the first and second). Stability may be somewhat reduced compared to the model of Fig. 7(a), but such reduction is likely to be minimal. Note that holes for equalization of pressure will exist in the first and second membranes and that only the third membrane (not directly exposed to the environment) constitutes the gas-tight seal.

Rigorous stability analysis of these structures appears to be more complex than that of the current model and has not yet been carried out.

## Acknowledgements

The author would like to acknowledge D. Van Winkle, S. Safron, F. Flaherty, E. Manousakis, L. Van Dommelen, S. Bellenot, and J. Skofronick for useful discussions, I. Winger for introducing polyester film to the project and for recognizing the project's application to "solar chimney" [3], R. Fatema for assistance in building the first model to successfully contain partial vacuum, and MARTECH for logistical support. Additionally the author would like to acknowledge the two anonymous reviewers who patiently worked through the manuscript to bring it into its present form.

## Nomenclature

$x_{ni}$ is the displacement of the $i^{\text{th}}$ degree of freedom in the $n^{\text{th}}$ unit cell.

$c$ and $\gamma$ are the pretensioning $T$ and elastic constant $K$ of one radial membrane respectively.

$a$ and $\alpha$ are the effective pretensioning $T$ and effective elastic constant $K$ of one segment of the inside tension hoop respectively.

$b$ and $\beta$ are the effective pretensioning $T$ and effective elastic constant $K$ of one segment of the outside tension hoop respectively.

$P$ is the absolute pressure in the pressurized regions.

$R$ is the vertex radius of the outside tension hoop.

$U$ is the potential energy of the system.

$N$ is the multiplicity of the axial symmetry. (The system has $N$-fold axial symmetry.)

For notational convenience we define $S \equiv R - 1$, $C \equiv \dfrac{c}{S}$, $A \equiv \dfrac{a}{2\delta}$, $B \equiv \dfrac{b}{2\delta R}$, $Q \equiv P - 1$, $\theta \equiv \dfrac{2\pi}{N}$,

$D \equiv \cos\dfrac{\theta}{2}$, $\delta \equiv \sin\dfrac{\theta}{2}$, $E \equiv \dfrac{\cos\theta}{2} = \dfrac{D^2 - \delta^2}{2}$, and $\varepsilon \equiv \dfrac{\sin\theta}{2} = D\delta$.

## Appendix A

The "reference" values discussed in section "A Note about Units" above are given below. Here $\Delta z$ is the differential length considered in the axial dimension.

$R_{\text{unit}} = x_{\text{unit}} = l_{\text{unit}} =$ radius to inside nodes

575  $P_{\text{unit}} = \Delta p_{\text{unit}} = \text{ambient pressure}$

$T_{\text{unit}} = a_{\text{unit}} = b_{\text{unit}} = c_{\text{unit}} = R_{\text{unit}} P_{\text{unit}} \Delta z$

$K_{\text{unit}} = \alpha_{\text{unit}} = \beta_{\text{unit}} = \gamma_{\text{unit}} = P_{\text{unit}} \Delta z$

$U_{\text{unit}} = R_{\text{unit}}{}^2 P_{\text{unit}} \Delta z$

580  **Appendix B**

As an example of (1), we calculate $\Delta U_{\text{membrane}}$ for the membrane in the outer tension hoop connecting the unit cells $n = 5$ and $n = 6$.

$$\Delta U_{\text{membrane}} = T(\Delta l) + \frac{1}{2} K (\Delta l)^2$$

585  $T = b = 2\delta B R$

$K = \beta$

$\Delta l = l - l_0$

$l_0 = 2R\delta$

$$l = \sqrt{\left( \begin{bmatrix} \delta \\ -D \\ \delta \\ D \end{bmatrix}^{\text{T}} \begin{bmatrix} x_{52} \\ x_{53} \\ x_{62} \\ x_{63} \end{bmatrix} + 2R\delta \right)^2 + \left( \begin{bmatrix} D \\ \delta \\ -D \\ \delta \end{bmatrix}^{\text{T}} \begin{bmatrix} x_{52} \\ x_{53} \\ x_{62} \\ x_{63} \end{bmatrix} \right)^2}$$

590  where the first and second terms in parenthesis are the displacements parallel and perpendicular to the membrane respectively. The square root is then evaluated to give

$$\Delta l = \begin{bmatrix} \delta \\ -D \\ \delta \\ D \end{bmatrix}^{\text{T}} \begin{bmatrix} x_{52} \\ x_{53} \\ x_{62} \\ x_{63} \end{bmatrix} + \begin{bmatrix} x_{52} \\ x_{53} \\ x_{62} \\ x_{63} \end{bmatrix}^{\text{T}} \frac{1}{4R\delta} \begin{bmatrix} D^2 & \varepsilon & -D^2 & \varepsilon \\ \varepsilon & \delta^2 & -\varepsilon & \delta^2 \\ -D^2 & -\varepsilon & D^2 & -\varepsilon \\ \varepsilon & \delta^2 & -\varepsilon & \delta^2 \end{bmatrix} \begin{bmatrix} x_{52} \\ x_{53} \\ x_{62} \\ x_{63} \end{bmatrix} + \cdots$$

where the ellipsis represents terms third-order in the $x_{ni}$. We then insert these into (1) and find that $\Delta U_{\text{membrane}}$ is equal to

595  $$\begin{bmatrix} 2BR\delta^2 \\ -2BR\varepsilon \\ 2BR\delta^2 \\ 2BR\varepsilon \end{bmatrix}^{\text{T}} \begin{bmatrix} x_{52} \\ x_{53} \\ x_{62} \\ x_{63} \end{bmatrix} + \frac{1}{2} \begin{bmatrix} x_{52} \\ x_{53} \\ x_{62} \\ x_{63} \end{bmatrix}^{\text{T}} \begin{bmatrix} BD^2 + \beta\delta^2 & B\varepsilon - \beta\varepsilon & -BD^2 + \beta\delta^2 & B\varepsilon + \beta\varepsilon \\ B\varepsilon - \beta\varepsilon & B\delta^2 + \beta D^2 & -B\varepsilon - \beta\varepsilon & B\delta^2 - \beta D^2 \\ -BD^2 + \beta\delta^2 & -B\varepsilon - \beta\varepsilon & BD^2 + \beta\delta^2 & -B\varepsilon + \beta\varepsilon \\ B\varepsilon + \beta\varepsilon & B\delta^2 - \beta D^2 & -B\varepsilon + \beta\varepsilon & B\delta^2 + \beta D^2 \end{bmatrix} \begin{bmatrix} x_{52} \\ x_{53} \\ x_{62} \\ x_{63} \end{bmatrix} + \cdots .$$

The elements of the first column vector contribute to $F_{ni}$ and the elements of the matrix contribute to $K_{nimj}$. For example, the contribution to $K_{5362}$ is $-B\varepsilon - \beta\varepsilon$.

**Appendix C**

600

As an example of (2), we calculate $\Delta U_{\text{gas}}$ for one sector of the volume inside the outer tension hoop. We take the triangular sector whose three vertices are the two nodes on the outer tension hoop in unit cells $n = 7$ and $n = 8$ and the center of the vacuum chamber.

$$\Delta U_{\text{gas}} = -\Delta p \cdot \Delta V$$

$$\Delta p = P - 1 = Q$$

$$\Delta V = V - V_0$$

$$V_0 = \varepsilon R^2$$

$$V = \frac{1}{2} \det \begin{bmatrix} DR + Dx_{72} + \delta x_{73} & -\delta R - \delta x_{72} + Dx_{73} \\ DR + Dx_{82} - \delta x_{83} & \delta R + \delta x_{82} + Dx_{83} \end{bmatrix}$$

$$\Delta U_{\text{gas}} = \begin{bmatrix} -\varepsilon QR \\ EQR \\ -\varepsilon QR \\ -EQR \end{bmatrix}^{\mathrm{T}} \begin{bmatrix} x_{72} \\ x_{73} \\ x_{82} \\ x_{83} \end{bmatrix} + \frac{1}{2} \begin{bmatrix} x_{72} \\ x_{73} \\ x_{82} \\ x_{83} \end{bmatrix}^{\mathrm{T}} Q \begin{bmatrix} 0 & 0 & -\varepsilon & -E \\ 0 & 0 & E & -\varepsilon \\ -\varepsilon & E & 0 & 0 \\ -E & -\varepsilon & 0 & 0 \end{bmatrix} \begin{bmatrix} x_{72} \\ x_{73} \\ x_{82} \\ x_{83} \end{bmatrix}$$

Again, the elements of the first column vector contribute to $F_{ni}$ and the elements of the matrix contribute to $K_{nimj}$. For example, the contribution to $F_{73}$ is $-EQR$. (Notice the sign reversal on $F_{ni}$ in (3).)